\documentclass[%
 reprint,
superscriptaddress,
 amsmath,amssymb,
 aps,
]{revtex4-1}

\usepackage{graphicx}% Include figure files
\usepackage{dcolumn}% Align table columns on decimal point
\usepackage{bm}% bold math
\usepackage{mathrsfs}
\begin{document}

\preprint{pra}

\title{QFT Treatment of the Klein Paradox}
%\thanks{A footnote to the article title}

\author{C. Xu}
 \email{xu.chuang.phy@gmail.com;}

\affiliation{School of Science, China University of Mining and Technology, Beijing 100083, China}

\author{Y. J. Li}
 \email{lyj@aphy.iphy.ac.cn.}

\affiliation{School of Science, China University of Mining and Technology, Beijing 100083, China}

\date{\today}

\begin{abstract}
It is well known that, Klein paradox is one of the most exotic and counterintuitive consequences of quantum theory. Nevertheless, many discussions about the Klein paradox are based upon single-particle Dirac equation in quantum mechanics rather than quantum field method. By using the path integral formalism, we evaluate the reflection and transmission coefficients up to the lowest order for the electron scattering by the finite square barrier potential. Within the context of assuming the step potential is the limiting case of the finite square barrier potential, we explain the Klein paradox that is caused by the ill-definition of the step potential.
\end{abstract}

\pacs{Valid PACS appear here}%PACS, the Physics and Astronomy
\maketitle

%\tableofcontents

\section{INTRODUCTION}

In 1929, soon after Paul Dirac proposing the new theory of the electron\cite{Dirac1928}, the Swedish physicist Oskar Klein, by applying the Dirac equation to treat the electron scattering from a step potential, obtained an astonishing result. It predicts that, seemingly, the low energy relativistic quantum electron can pass through a sufficiently high potential without the exponential damping expected in usual quantum tunneling processes. Furthermore, in pace with the strength of potential approaches infinity, the incoming electron almost pass through the potential transparently. This phenomenon, was known as the Klein paradox\cite{Klein1929}.

There have been various attempts to interpret this result since the publication of Klein's paper, some detailed discussions can be found in Refs.\ \cite{Bjorken1964,Greiner1985,Dombey1999,Merad2000,Dombey2000,Hejcik2008,Gaume2012}. Its earlier interpretations are by Sauter\cite{Sauter1931}, Heisenberg and Euler\cite{Heisenberg1936}. In 1941, Hund firstly, utilizing theoretical method of second quantization for the scalar field, demonstrated that if the potential is strong enough, a non-zero constant electric current, given by an integral over the transmission coefficient with respect to energy, has to be present\cite{Hund1941}. Ten years later, Schwinger used the non-perturbational method, creatively, to link the pair-creation probability with the imaginary part of the effective Lagrangian density\cite{Schwinger1951}. As the Dirac equation is actually regarded as the equation of the spinor field rather than scalar field, in 1981, Hansen and Ravndal, using the standard operator method of quantum field theory, generalized Hund's work to the spinor field\cite{Hansen1981}.

As quantum field theory has achieved unprecedented success, we have reason to believe that, it is reliable to treat the Klein paradox by using path integral method. In this work, without associating the process of electron-positron pair creation, we give an another possible interpretation for producing the Klein paradox. That is, the Klein paradox is caused by the ill-definition of the step potential. The paper is organized as follows. In Sec. II we review the framework of path integral treatment for the electron scattering by electromagnetic interaction. In Sec. III, following the Klein's initial ideal, we use the path integral formalism to calculate the reflection and transmission coefficients for the electron in the present of the finite square barrier potential. Then we generalize the conclusion of the finite square barrier potential to the case of the step potential. In the last section we give a brief summary.

\section{Path Integral Treatment}
In this section, using path integral method, we can give a general result for the electron scattering by a time-independent electromagnetic background field. Indeed, according to the general principle of quantum field theory, we need to quantize the spinor field and electromagnetic field simultaneously. However, in order to compare our final result directly with those of the single-particle Dirac equation, we shall not quantize the electromagnetic field. Instead, treat the field as a given, classical four-dimensional potential $A_\mu(x)$ throughtout. Besides, unless specified otherwise, we always use the natural units and work in the interaction picture.

The fundamental object of path integral formalism is the generating functional of correlation functions\cite{Peskin1995}. For the spinor field, it reads
\begin{align}
 Z[\bar{\eta},\eta]=&\int\mathscr{D}\bar{\psi}\mathscr{D}\psi\ \textrm{exp}
\{iS_0[\bar{\psi},\psi]+iS_{int}[\bar{\psi},\psi]\nonumber\\
&+iS_{source}[\bar{\psi},\psi,\bar{\eta},\eta]\},
\end{align}
where $S_0$, $S_{int}$ and $S_{source}$ are free action, interactional action and source action respectively, $\psi$ and $\eta$ are Grassmann fields whose values are anticommuting numbers.

First, we need to refer the definition of the function derivatives, $\delta/\delta\eta$ and $\delta/\delta\bar{\eta}$, as follow. In the case of the four dimensions, the functional derivatives obey the basic axiom
\begin{align}
&\frac{\delta}{\delta\eta(x)}\eta(y)=\delta^{(4)}(x-y), \nonumber\\
&\frac{\delta}{\delta\bar{\eta}(x)}\bar{\eta}(y)=\delta^{(4)}(x-y).
\end{align}
In addition, if $\eta$ and $\bar{\eta}$ are Grassmann numbers, they satisfy
\begin{equation}
 \frac{d}{d\eta}\bar{\eta}\eta=-\frac{d}{d\eta}\eta\bar{\eta}=-\bar{\eta}.
\end{equation}

To evaluate the path integral $Z[\bar{\eta},\eta]$ more generally, we must split up the exponential into interaction and free term including source. Meanwhile, to fix the problem, we can use a trick that replaces the fields $\bar{\psi}$ and $\psi$ in the interaction part by functional derivatives. Then, the Eq. (1) becomes
\begin{align}
Z[\bar{\eta},\eta]=
&\int\mathscr{D}\bar{\psi}\mathscr{D}\psi\ \textrm{exp}\{iS_{int}[\bar{\psi},\psi]\}
\textrm{exp}\{iS_0[\bar{\psi},\psi]\nonumber\\
&+iS_{source}[\bar{\psi},\psi,\bar{\eta},\eta]\}\nonumber\\
=&\int\mathscr{D}\bar{\psi}\mathscr{D}\psi\ \textrm{exp}\{iS_{int}[+i\delta/\delta\eta,-i\delta/\delta\bar{\eta}]\}\nonumber\\
&\textrm{exp}\{iS_0[\bar{\psi},\psi]
+iS_{source}[\bar{\psi},\psi,\bar{\eta},\eta]\}.
\end{align}
It would be nice if we can pull the interaction part out of the integral and perform the remaining integral. So we have
\begin{equation}
 \begin{split}
Z[\bar{\eta},\eta]=\textrm{exp}\Big\{iS_{int}
\Big[+i\frac{\delta}{\delta\eta},-i\frac{\delta}{\delta\bar{\eta}}\Big]\Big\}Z_0[\bar{\eta},\eta],
 \end{split}
\end{equation}
where $Z_0[\bar{\eta},\eta]$ is the generating functional of the free spinor field
\begin{align}
&Z_0[\bar{\eta},\eta]\nonumber\\
=&\int\mathscr{D}\bar{\psi}\mathscr{D}\psi\ \textrm{exp}
\{iS_0[\bar{\psi},\psi]
+iS_{source}[\bar{\psi},\psi,\bar{\eta},\eta]\}\nonumber\\
=&\int\mathscr{D}\bar{\psi}\mathscr{D}\psi\ \textrm{exp}
\Big\{i\int d^4x[\bar{\psi}(i\partial \!\!\!/-m)\psi+\eta\bar{\psi}+\psi\bar{\eta}]\Big\}\nonumber\\
=&Z_0[0,0]\cdot\textrm{exp}\Big\{-\int d^4xd^4y\bar{\eta}(x)S_F(x-y)\eta(y)\Big\}.
\end{align}
Notice that $Z_0[0,0]$ actually is the Gaussian integral and can be performed precisely. So as to simply the subsequent calculations, we will drop it directly. Therefore, we redefines it as
\begin{equation}
 Z_0[\bar{\eta},\eta]=\textrm{exp}\Big\{-\int d^4xd^4y\bar{\eta}(x)S_F(x-y)\eta(y)\Big\}.
\end{equation}
where $S_F(x-y)$ is called the Feynman propagator for the spinor field, it is to write
\begin{equation}
 S_F(x-y)=\int\frac{d^4p}{(2\pi)^4}\frac{i(p\!\!\!/+m)}{p^2-m^2+i\epsilon}
e^{-p\cdot(x-y)}.
\end{equation}

In the cause of describing the interaction of spinor field and electromagnetic field, we choose $S_{int}[\bar{\psi},\psi]$ as
\begin{align}
S_{int}[\bar{\psi},\psi]&=\int d^4z\mathscr{L}_{int}[\bar{\psi}(z),\psi(z)]\nonumber\\
&=-e\int d^4z A_\mu(z)\bar{\psi}(z)\gamma^\mu\psi(z).
\end{align}

On the basis of path integral formulism, the two-point correlation function for the spinor field
is given by
\begin{align}
&\langle\Omega|T\psi_H(x_1)\bar{\psi}_H(x_2)|\Omega\rangle\nonumber\\
=&\frac{\int\mathscr{D}\bar{\psi}\mathscr{D}\psi\ \psi(x_1)\bar{\psi}(x_2) \textrm{exp}
\{iS[\bar{\psi},\psi]\}}
{\int\mathscr{D}\bar{\psi}\mathscr{D}\psi \textrm{exp}\{iS[\bar{\psi},\psi]\}}\nonumber\\
=&\frac{(-i\delta/\delta\bar{\eta}(x_1))(+i\delta/\delta\eta(x_2))
Z[\bar{\eta},\eta]}
{Z[\bar{\eta},\eta]}\Big|_{\bar{\eta},\eta=0},
\end{align}
where $|\Omega\rangle$ denotes the vacuum state of interacting  theory. $T$ is called time-ordered product, which instructs us to place the operators that follow in order with the latest to the left. And the subscript $H$ means working in the Heisenberg picture.

Consider Eq. (10), in fact, the disconnected diagrams in the numerator can be just canceled by the denominator, only the connected diagrams can make contributions to the correlation function. In other words, for the two-point correlation function, we can simply sum of all connected diagrams with two external points. Now it reads
\begin{align}
&\langle\Omega|T\psi_H(x_1)\bar{\psi}_H(x_2)|\Omega\rangle\nonumber\\
=&\Big(-i\frac{\delta}{\delta\bar{\eta}(x_1)}\Big)\Big(+i\frac{\delta}{\delta\eta(x_2)}\Big)
Z[\bar{\eta},\eta]\Big|^{connected}_{\bar{\eta},\eta=0}\nonumber\\
=&\frac{\delta}{\delta\bar{\eta}(x_1)}\frac{\delta}{\delta\eta(x_2)}\textrm{exp}\Big\{iS_{int}
\Big[+i\frac{\delta}{\delta\eta},-i\frac{\delta}{\delta\bar{\eta}}\Big]\Big\}\nonumber\\
&Z_0[\bar{\eta},\eta]\Big|^{connected}_{\bar{\eta},\eta=0}.
\end{align}

To compute $\langle\Omega|T\psi(x_1)\bar{\psi}(x_2)|\Omega\rangle$, we can expand the generating functional $Z[\bar{\eta},\eta]$ perturbatively in powers of $S_{int}$,
\begin{align}
&Z[\bar{\eta},\eta]\nonumber\\=&\Big(1+iS_{int}
\Big[+i\frac{\delta}{\delta\eta},-i\frac{\delta}{\delta\bar{\eta}}\Big]\nonumber\\
&-\frac{1}{2}S_{int}
\Big[+i\frac{\delta}{\delta\eta},-i\frac{\delta}{\delta\bar{\eta}}\Big]^2+\cdots\Big)
Z_0[\bar{\eta},\eta].
\end{align}

In order to obtain the reflection and transmission coefficients defined in quantum mechanics, for simplicity, we calculate a few non-trivial order contributions. The zero-order term in the expansion of (11) is given by
\begin{align}
&\frac{\delta}{\delta\bar{\eta}(x_1)}\frac{\delta}{\delta\eta(x_2)}
Z_0[\bar{\eta},\eta]\Big|^{connected}_{\bar{\eta},\eta=0}\nonumber\\
=&\frac{\delta}{\delta\bar{\eta}(x_1)}\Big[\int d^4x\bar{\eta}(x)S_F(x-x_2)\Big]Z_0[\bar{\eta},\eta]\Big|^{connected}_{\bar{\eta},\eta=0}\nonumber\\
=&S_F(x_1-x_2).
\end{align}
As we will see later, the zero-order term is just corresponding to the identity part of the S-matirx.

For the first-order term, it becomes complicated, we must calculate the quantity
\begin{align}
&\frac{\delta}{\delta\bar{\eta}(x_1)}\frac{\delta}{\delta\eta(x_2)}(iS_{int}[\bar{\psi},\psi])
Z_0[\bar{\eta},\eta]\Big|^{connected}_{\bar{\eta},\eta=0}\nonumber\\
=&-ie\int d^4 zA_\mu(z)\frac{\delta}{\delta\bar{\eta}(x_1)}\frac{\delta}{\delta\eta(x_2)}
\bar{\psi}(z)\gamma^\mu\psi(z)\nonumber\\
&Z_0[\bar{\eta},\eta]\Big|^{connected}_{\bar{\eta},\eta=0}.
\end{align}
Repeat to use the above trick, which is replacing the fields $\bar{\psi}$ and $\psi$ by functional derivatives. Thus, Eq. (14) becomes
\begin{align}
&-ie\int d^4 zA_\mu(z)\frac{\delta}{\delta\bar{\eta}(x_1)}\frac{\delta}{\delta\eta(x_2)}
\Big(+i\frac{\delta}{\delta\eta(z)}\Big)\nonumber\\
&\ \gamma^\mu\Big(-i\frac{\delta}{\delta\bar{\eta}(z)}\Big)
Z_0[\bar{\eta},\eta]\Big|^{connected}_{\bar{\eta},\eta=0}\nonumber\\
=&-ie\int d^4 zA_\mu(z)\gamma^\mu\frac{\delta}{\delta\bar{\eta}(x_1)}\frac{\delta}{\delta\eta(x_2)}
\frac{\delta}{\delta\eta(z)}\nonumber\\
&\ \frac{\delta}{\delta\bar{\eta}(z)}
Z_0[\bar{\eta},\eta]\Big|^{connected}_{\bar{\eta},\eta=0}\nonumber\\
=&-ie\int d^4 zA_\mu(z)\gamma^\mu\frac{\delta}{\delta\bar{\eta}(x_1)}\frac{\delta}{\delta\eta(x_2)}
\frac{\delta}{\delta\eta(z)}\frac{\delta}{\delta\bar{\eta}(z)}\nonumber\\
&\ \textrm{exp}\Big\{-\int d^4xd^4y\bar{\eta}(x)S_F(x-y)\eta(y)\Big\}
\Big|^{connected}_{\bar{\eta},\eta=0}\nonumber\\
=&ie\int d^4 zA_\mu(z)S_F(x_1-z)\gamma^\mu S_F(z-x_2).
\end{align}

Our next task is to establish the connection between the two-point correlation function and the S-matrix. Using the generalized Lehmann-Symanzik-Zimmermann reduction formula for the spinor field\cite{Lehmann1955}, we obtain
\begin{align}
&(\sqrt Z)^2\big\langle\mathbf{p}\big|S\big|\mathbf{k}\big\rangle\nonumber\\
=&\int d^4x_1e^{ip\cdot x_1}\int d^4x_2e^{-ik\cdot x_2}\bar{u}(p)(-i)(i{\partial \!\!\!/}_{x_1}-m)\nonumber\\
&\langle\Omega|T\psi_H(x_1)\bar{\psi}_H(x_2)|\Omega\rangle(-i)
(-i\loarrow{\partial \!\!\!/}_{x_2}-m)u(k).
\end{align}
where $u$ is the spinor wave function of the electron, $\bar{u}$ is the Dirac conjugation of $u$, and $Z$ is a c-number, known as wave function renormalization. Since we only work up to the first-order term, we set $Z=1$.

To isolate the interaction part of the S-matrix, we define the T-matrix by
\begin{equation}
     S=\mathbf{1}+iT.
\end{equation}
For the identity matrix, it distinctly indicates that,
\begin{equation}
     \mathscr{R}^{(0)}=0,\quad \mathscr{T}^{(0)}=1.
\end{equation}
where $\mathscr{R}^{(0)}$ and $\mathscr{T}^{(0)}$ are reflection and transmission amplitudes
at the level of the zero-order term.

The lowest non-trivial order contribution for T-matrix is the first-order term. So,
combine the result of (16), the only thing we need to do is to calculate the quantity
\begin{align}
&ie\int d^4x_1e^{ip\cdot x_1}\int d^4x_2e^{-ik\cdot x_2}
\int d^4 zA_\mu(z) \nonumber\\
&\bar{u}(p)(-i)(i{\partial \!\!\!/}_{x_1}-m)S_F(x_1-z)\gamma^\mu\nonumber\\& S_F(z-x_2)(-i)
(-i\loarrow{\partial \!\!\!/}_{x_2}-m)u(k)\nonumber\\
=&-ie\int d^4x_1e^{ip\cdot x_1}\int d^4x_2e^{-ik\cdot x_2}
\int d^4 zA_\mu(z)\nonumber\\
&\ \bar{u}(p)\delta^4(x_1-z)\gamma^\mu\delta^4(z-x_2)u(k)\nonumber\\
=&-ie\bar{u}(p)\gamma^\mu u(k)\int d^4 z e^{iz\cdot(p-k)}A_\mu(z).
\end{align}
Substitute the above result into Eq. (16), we have
\begin{equation}
\big\langle\mathbf{p}\big|iT\big|\mathbf{k}\big\rangle
=-ie\bar{u}(p)\gamma^\mu u(k)\cdot\tilde{A}_\mu(p-k),
\end{equation}
where $\tilde{A}_\mu(p-k)$ is the four-dimensional Fourier transform of $A_\mu(z)$,
\begin{equation}
\tilde{A}_\mu(p-k)=\int d^4 z e^{iz\cdot(p-k)}A_\mu(z).
\end{equation}

\section{Explicit Calculation For the barrier potential}
In this section, to link the result (20) to the Klien paradox, we consider the one-dimensional case. Suppose the electron travels along $z^3$ axis, let us set
\begin{align}
&k^\mu=(k^0,0,0,k^3)\nonumber\\
&p^\mu=(p^0,0,0,p^3)\nonumber\\
&A_\mu=(A_0(z^3),0,0,0),
\end{align}
where $k^0=\sqrt{(k^3)^2+m^2}$ and $p^0=\sqrt{(p^3)^2+m^2}$.
Then, the T-matrix becomes
\begin{align}
&\big\langle\mathbf{p}\big|iT\big|\mathbf{k}\big\rangle\nonumber\\
=&-ie\bar{u}(p)\gamma^0 u(k)\int d^2 z e^{iz\cdot(p-k)}A_0(z)\nonumber\\
=&-ieu^\dag(p^3)u(k^3)\int dz^0 e^{iz^0(p^0-k^0)}\int \nonumber\\&\ dz^3e^{-iz^3(p^3-k^3)}A_0(z^3)\nonumber\\
=&-iu^\dag(p^3)u(k^3)(2\pi)\delta(p^0-k^0)\int \nonumber\\&\ dz^3e^{-iz^3(p^3-k^3)}V(z^3).
\end{align}
In the last line, we define $V(z^3)=eA_0(z^3)$.

To describe the electron scattering process completely, we have to integrate over the final momentum $p^3$ with the Lorentz-invariant measure,
\begin{align}
&\int\frac{d^3p}{(2\pi)^3}\frac{1}{2E(\mathbf{p})}\big\langle
\mathbf{p}\big|iT\big|\mathbf{k}\big\rangle \nonumber\\
=&-i\int\frac{dp^3}{(2\pi)^3}\frac{1}{2E(p^3)}u^\dag(p^3)u(k^3)\int dz^0 \nonumber\\&\ e^{iz^0(p^0-k^0)}\int  dz^3e^{-iz^3(p^3-k^3)}V(z^3)\nonumber\\
=&-i\int\frac{dp^3}{(2\pi)^2}\frac{1}{2E(p^3)}u^\dag(p^3)u(k^3)\delta(p^0-k^0)\nonumber\\&\int dz^3e^{-iz^3(p^3-k^3)}V(z^3).
\end{align}
With the aim of performing the integration over $p^3$, we use the identity\cite{Leo2009}
\begin{equation}
\delta(p^0-k^0)=\frac{k^0}{k^3}[\delta(p^3-k^3)+\delta(p^3+k^3)].
\end{equation}
Then, Eq. (24) will be split into two separate parts. One is for transmission and another is for reflection.

For the purpose of explaining the Klein paradox, we choose $V(z^3)$ as a barrier potential
\begin{eqnarray*}
V(z^3)=
\begin{cases}
V            & 0 < z^3 < L, \\
0            & z^3 < 0, z^3 > L.
\end{cases}
\end{eqnarray*}

\begin{figure}[!thb]
\includegraphics[width=8.7cm]{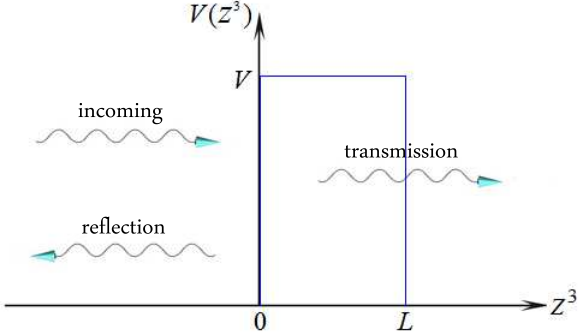}
\caption{\label{fig:epsart} The electron travels along with $z^3$ axis and be scattered by one-dimensional barrier potential.}
\end{figure}

Consequently, for reflection, $p^3=-k^3$, we have
\begin{align}
\mathscr{R}^{(1)}
=&-i\int\frac{dp^3}{(2\pi)^2}\frac{1}{2E(p^3)}u^\dag(p^3)u(k^3)\frac{k^0}{k^3}\nonumber\\
&\ \delta(p^3+k^3)\int dz^3e^{-iz^3(p^3-k^3)}V(z^3)\nonumber\\
=&-i\frac{u^\dag(-k^3)u(k^3)}{(2\pi)^2k^3}\int dz^3e^{i2z^3k^3)}V(z^3)\nonumber\\
=&-i\frac{m}{k^3}\int dz^3e^{i2z^3k^3)}V(z^3)\nonumber\\
=&-i\frac{mV}{(k^3)^2}\sin(k^3L)e^{ik^3L}.
\end{align}
For transmission, $p^3=k^3$, we have
\begin{align}
\mathscr{T}^{(1)}
=&-i\int\frac{dp^3}{(2\pi)^2}\frac{1}{2E(p^3)}u^\dag(p^3)u(k^3)\frac{k^0}{k^3}\nonumber\\
&\ \delta(p^3-k^3)\int dz^3e^{-iz^3(p^3-k^3)}V(z^3)\nonumber\\
=&-i\frac{u^\dag(k^3)u(k^3)}{(2\pi)^2k^3}\int dz^3V(z^3)\nonumber\\
=&-i\frac{k^0}{k^3}VL.
\end{align}

Thus, up to the first-order term, we finally obtain the reflection and transmission coefficients
\begin{align}
  \nonumber R=&\big|\mathscr{R}\big|^2=\big|\mathscr{R}^{(0)}+\mathscr{R}^{(1)}\big|^2\\
  \nonumber  =&\Big|-i\frac{mV}{(k^3)^2}\sin(k^3L)e^{ik^3L}\Big|^2\\
             =&\frac{m^2V^2}{(k^3)^4}\sin^2(k^3L),
\end{align}
and
\begin{align}
  \nonumber T=&\big|\mathscr{T}\big|^2=\big|\mathscr{T}^{(0)}+\mathscr{T}^{(1)}\big|^2\\
  \nonumber  =&\Big|1-i\frac{k^0}{k^3}VL\Big|^2\\
  \nonumber  =&1+(k^0)^2V^2L^2/(k^3)^2\\
             =&1+[1+m^2/(k^3)^2]V^2L^2.
\end{align}
Here we use the relation $k^0=\sqrt{(k^3)^2+m^2}$ again.

The results (28) and (29) are in keeping with our participation. Notice that, although the result (29) seems greater than one, this is because of our calculation just up to the first order. If we cite the result in Refs.\ \cite{Leo2009}, which is related to the second order, the transmission coefficient will less than one. Furthermore, in fact, in order to ensure the positive definiteness of transmission coefficient, namely $T\geq 0$, we must restrict the length of barrier potential. In other words, for given $k^3$ and $V$, it is unreasonable to let $L$ approach infinity. Meanwhile, it's not difficult to find that we still obtain the nonphysical result if we replace to use the step potential.

\section{CONCLUSIONS}
Since Schwinger's work, it becomes clear that the vacuum of quantum electrodynamics is unstable against the electron-positron pair creation in the present of external electromagnetic field\cite{Brezin1970,Narozhny1972,Marinov1977,Casher1979,Kim2006}. Accordingly, many subsequent papers adopt the view of spontaneous pair creation to discuss the Klein paradox. Whereas, according to Schwinger's prediction and energy conservation, we need at least $V\sim2m_e$ electric field strength to produce the electron-positron pairs. On the other hand, we know the so-called condition of occurring Klein paradox that is when the energies of incoming electrons satisfy $m_e<E<V-m_e$, the transmission will be into classically forbidden region and more electrons reflected than incident. That is to say, to produce the apparent paradox, the potential needs to satisfy $V>E+m_e$. Generally, as the incoming electrons have definite kinetic energies, the critical value of producing the Klein paradox may be great than pair creation. It is possible to appear the embarrassing situation that is the potential is not enough strong to produce the Klein paradox, but the electron-positron pair creation process has taken place.

Through our calculations, we demonstrate that the main reason induced the Klein paradox is the step potential itself. It is not necessarily connected to pair-creation process. Physically, in the context of quantum field theory description, the step potential is ill-defined. An infinite length of the step potential is an unrealistic idealization. If we treat the finite square barrier potential, in the Klein zone, we find instead no paradox. This is the natural result of path integral treatment. As we know, quantum mechanics never imposes the restrictions on the specific form of interactions. Nevertheless, rather than quantum mechanics, in the spirit of quantum filed theory, the form of interactions are required.

\begin{acknowledgments}
We thank Y. F. Xu, J. Chen and F. Feng for several helpful discussions. We acknowledge support from the National Natural Science Foundation of China Grant No. 11374360 and No. 11405266 and National Basic Research Program of China Grant No. 2013CBA01504.
\end{acknowledgments}

\nocite{*}

\bibliography{aps}

\end{document}